# Boosting XML Filtering with a Scalable FPGA-based Architecture

Abhishek Mitra, Marcos R. Vieira, Petko Bakalov, Walid Najjar, Vassilis J. Tsotras
University of California
Riverside, CA 92521, USA
{amitra,mvieira,pbakalov,najjar,tsotras}@cs.ucr.edu

## ABSTRACT
The growing amount of XML encoded data exchanged over the Internet increases the importance of XML based publish-subscribe (pub-sub) and content based routing systems. The input in such systems typically consists of a stream of XML documents and a set of user subscriptions expressed as XML queries. The pub-sub system then filters the published documents and passes them to the subscribers. Pub-sub systems are characterized by very high input ratios, therefore the processing time is critical. In this paper we propose a "pure hardware" based solution, which utilizes XPath query blocks on FPGA to solve the filtering problem. By utilizing the high throughput that an FPGA provides for parallel processing, our approach achieves drastically better throughput than the existing software or mixed (hardware/software) architectures. The XPath queries (subscriptions) are translated to regular expressions which are then mapped to FPGA devices. By introducing stacks within the FPGA we are able to express and process a wide range of path queries very efficiently, on a scalable environment. Moreover, the fact that the parser and the filter processing are performed on the same FPGA chip, eliminates expensive communication costs (that a multi-core system would need) thus enabling very fast and efficient pipelining. Our experimental evaluation reveals more than one order of magnitude improvement compared to traditional pub/sub systems.

## 1. INTRODUCTION
Publish/subscribe applications (or simply pub-sub) are an important class of content-based dissemination systems where the message transmission is guided by the message content, rather than its destination IP address. System architectures may vary (centralized within a server or distributed over a network of brokers) but they all follow the same asynchronous event-based communication paradigm. The input is a stream of messages, generated outside of the system by a set of *publishers*. These messages are then selectively delivered to interested *subscribers* that have declared their interest by submitting *profiles* to the pub-sub system. This process is also known as *message filtering*. Examples of pub-sub systems include notification websites (e.g. *www.hotwire.com*, *news.google.com* and *www.ticketmaster.com*), where a user can subscribe for specific events ("Rock concerts in Chicago") and get automatic notifications when the event occurs. Increasingly such environments are becoming XML-based, i.e., the messages are exchanged in XML while the users express their subscriptions using XML query languages like XPath [31].

Given the high volumes of messages and profiles, the filtering process becomes a critical performance requirement for pub-sub systems. The predominant solutions to this problem perform clustering of the user profiles based on their similarity in order to narrow down the search in the profile space. This is done by the use of Finite State Machines (FSM). In particular, elements of the user profiles are mapped to states in the state machine. The clustering is then performed by combining multiple profiles in a single FSM by analyzing and discovering the common profile paths. Since user profiles are typically known in advance (i.e., profiles play the role of data, while documents play the role of traditional queries) it is possible to be analyzed and clustered as needed before the filtering process starts.

When a document arrives in a pub-sub system, it is parsed by an event-driven parser like SAX [1] that reports low level parsing events such as: "start document", "start element", etc. As events are produced by the SAX parser, they are processed by the filtering system which uses them to drive transitions between the FSM states. For example, a transition is taken from the current FSM state if there is an outgoing edge labeled with the tag currently being processed. If during this process an "accept" FSM state is reached the document satisfies the corresponding profile(s) associated with that state.

Implementing the above approach on a traditional von Neumann architecture would requires multiple clock cycles per instruction. Consider for example, the "high level" task of identifying an "open" tag during parsing. This corresponds to multiple low level instructions (e.g. *load* and *store*) where the execution of every such instruction requires at least one clock cycle. This issue is known as the *von Neumann bottleneck* and can limit the filtering speed to few hundreds of clock cycles per processing a single XML tag.

Given the above bottleneck of von Neumann machines, an attempt to improve performance is to execute the tasks in parallel by adding more resources (i.e., many processors). While this idea will not eliminate the bottleneck (each processor still uses multiple clock cycles per operation) it will also create a large communication overhead between the processors. For example, one could pipeline the parsing with the filtering tasks by running them on different processors. However, when the parser produces an event it needs to notify (communicate) the filtering processor about this event (thus creating large interprocessor communication cost).

The way to resolve this limitation is to use a non-traditional highly parallel architecture. In this paper we present a novel filtering approach which is based on the use of Field-Programmable Gate Arrays (FPGA).

FPGAs are increasingly being made available as co-processors on high-performance computation systems. They are packaged in modules, which are dropped in CPU sockets on server motherboards with bridges to the FSB / Quickpath [27] [36] links on Intel platforms and Hypertransport [6] link on AMD platforms. High density FPGAs such as Xilinx Virtex-4LX 200 [37] [38] and Altera Stratix EP2S80F [2] have millions of logic gates, abundant high speed dual port memory, ALU blocks on the silicon fabric, and have high speed multi Gbps speed I/O ports [38] [3] . These high density FPGAs can be used to implement in hardware the computationally intensive portions of the software code. Multithreaded software components with streaming data input and output like the pub-sub applications are ideal candidates for acceleration on FPGA co-processing systems since a huge amount of data can be processed in parallel on the FPGA.

Since pub-sub XML filtering involves multiple queries processed over a single document data-stream, it is possible to utilize FPGAs for parallelizing the filtering performance. Each query can be implemented on the FPGA unit as a hardware datapath circuit and with appropriate optimizations it is possible to fit thousands of queries on a single FPGA chip. Moreover, having the parallel processing modules implemented on the same chip eliminates the need for expensive communications between them. This in turn allows for full pipelining of the parsing and filtering processes: as an event is produced by the parser it is immediately forwarded to the filtering module. This results in accelerated query processing and furthermore leads to substantial savings in a general purpose computation infrastructure by reducing the amount of power required by the CPUs.

In this paper we present a "proof of concept" for the use of FPGAs in boosting XML filtering performance. We utilize a four step approach that converts such query into hardware description, suitable for implementation on FPGA. The first step involves conversion of an XPath query to PERL compatible regular expressions (PCREs). The regular expressions are clustered by their common prefixes in order to produce more compact representation on the board and are then translated to VHDL using our "regex to VHDL" compiler [24]. Moreover, in order to support parent-child relationships, we introduce the use of stacks and modify the regular expression hardware to use them. The highly optimized VHDL code is then deployed on the FPGA board. The stream of documents is forwarded to the board where it is processed with high degree of parallelism. Our experimental evaluation reveals that this architecture achieves orders of magnitude improvement in the terms of running time compared to the state of the art software based XML filtering systems.

The paper is organized as follows: Section 2 presents related work. Section 3 provides in depth description of the proposed architecture. Section 4 presents an experimental evaluation of the FPGA approach compared to the state of the art software counterparts. Finally conclusions and open problems for further research appear in section 5.

## 2. RELATED WORK

One of the first works that addressed XML filtering is the XFilter [4]. This approach defines a Finite State Machine (FSM) for each XPath user profile. Every tag (element) in the profile becomes a state in the FSM, while the last tag becomes the accept state in that FSM. These machines are then executed concurrently for each message in the input. In particular, a 'start element' event drives the machine through its various transitions from state to state, while an 'end element' event makes a transition backward to a previous state. Finally, if an accepted state is reached, the document is reported as a match to the corresponding profile's subscriber. Later, the YFilter [11] system improved the matching performance by combining all profiles into a single Nondeterministic Finite Automaton (NFA). Common profile prefixes are combined and represented with a single set of states. This allows dramatic reduction in the number of states needed to represent the set of user profiles. It also improves the filtering performance of the system by processing common profile paths only once.

Other FSM-based approaches use different techniques for building the state machine as well as different types of machines. For example, [14] builds a single deterministic push down automaton using a lazy approach, [12] employs a lazily built Deterministic Finite Automata (DFA), [22] builds a transducer, which employs a DFA with a set of buffers, and [28] employs a hierarchical organization of push down transducers with buffers.

All these solutions are similar in the sense that they traverse the provided input document in a top-down fashion (i.e. in-order traversal) while advancing the state machine for each XML element (or attribute) read. Another proposed approach is to use a bottom-up traversal of the document. This idea takes into consideration the fact that an XML document typically has its more selective elements located at its leaves and uses them to perform early pruning in the query space. Examples of systems which utilize the bottom-up approach include FiST [16] and BUFF [25].

The NFA based approaches discussed above are entirely *software* based solutions using the standard *von Neumann organization*. None of them takes advantage of specialized architectures to overcome the bottleneck problem which appears during XML document filtering.

Previous works [23, 19, 33] that have used FPGAs for processing XML documents have mainly dealt with the problem of XML parsing which in turn is transformed to implementing regular expressions on FPGAs. In particular, [23] proposes the *ZuXA* engine to parse XML documents. This engine employs state machines for efficient parsing based on set of rules. The paper however does not provide any discussion how this engine can be adapted to work with the XPath or twig profiles common in the pub-sub systems. A related FPGA based regular expression language parser adapted for content based routing of an XML stream has been demonstrated in [26].

There is also a large amount of research related to implementing regular expressions on FPGAs [32, 18]. Here we build on our previous works [24] where we compiled PERL Compatible Regular Expressions (PCRE) to VHDL for accelerating intrusion-detection system rules using FPGAs. However, XPath query evaluation is more complex than plain regular expressions. To this end we introduce appropriate stacks that are implemented on the FPGA device.

The works in [33, 19] propose the use of a mixed hardware/software architecture to solve simple XPath queries having only parent-child axis. A finite state machine implemented in FPGAs is facilitated to parse the XML document and to provide partial evaluation of XPath predicates. The results are then reported to the software part for further processing. Similarly to the *ZuXA* engine, this architecture can only support simple XPath queries with only parent-child axis.

There are also approaches that use specialized parallel architectures for XML processing [17, 20, 21]. In particular, [17] uses the Cell Broadband Engine multi-processor which consists of 8 independent processors (SPEs) that run the same software. This approach achieves parallelism by parsing (eight) XML documents in parallel at a time. Each processor implements the FSM of the *ZuXA* engine [23]. In addition to be only suitable for XML parsing, this solution is a combination of hardware-software approach. Similarly, the work in [20, 21] addresses ways to load-balance parallel threads for low-level XML processing (e.g., XML parsing). There is also work on running XML queries over documents that are fragmented among many processors [8, 10] and achieving parallelism through partial query evaluation; nevertheless, this is an orthogonal problem to filtering.

To the best of our knowledge our system is the first one to provide an *entirely* hardware solution to the XML filtering problem in pub-sub systems. It is also the first one able to efficiently evaluate complex XPath queries with different types of navigation directions (parent-child "/" as well as ancestor-descendant "//" axis) over the stream of XML documents. While parallelism can be achieved with multi-core machines (as a software-hardware solution), FPGAs offer a viable alternative due to their power efficiency (less power consumption and cooling costs) [34, 15] as well as higher throughput. The work in [13], quantitatively demonstrates the benefits of using FPGAs over general purpose CPUs for streaming applications. While multi-core systems come with 2 and 4 CPUs it is not always feasible to achieve proportional speed-up due to the bottleneck in shared cache memory and the front side bus.

## 3. IMPLEMENTING XPATH PROFILES ON FPGAS

We start this section with a short description of the FPGA architecture and the properties that make it appealing for XML filtering. This is followed by a general overview (Figure 2) of our compilation workflow, which loads the filtering logic on the FPGA chip. Finally we present a detailed description of the individual steps in the workflow; this includes two optimizations, namely the common prefix optimization and the area efficient character decoder.

A Field-Programmable Gate Array (or FPGA) is a semiconductor device containing programmable logic components termed "Configurable Logic Blocks" (CLB) connected trough programmable interconnections. An illustration of a typical FPGA architecture appears in Figure 1. The interconnections inside the device allow logic blocks to be interconnected as needed by the user in order to implement specific logic. Such devices allow implementation of multiple datapaths operating in parallel which makes them suitable for streaming applications like XML parsing and filtering. Moreover, because the datapath is implemented in hardware, the load and store operations from the von Neumann model are eliminated resulting in more efficient processing.

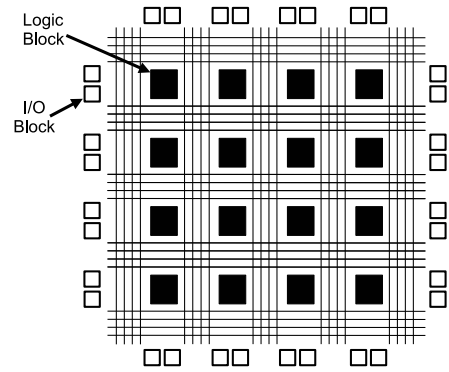

Figure 1: General Architecture of an FPGA. The reconfigurable hardware is realized with programmable SRAM blocks, called CLB (Configurable Logic Blocks) and programmable routing interconnects. A bitstream can program an FPGA to realize the required hardware.

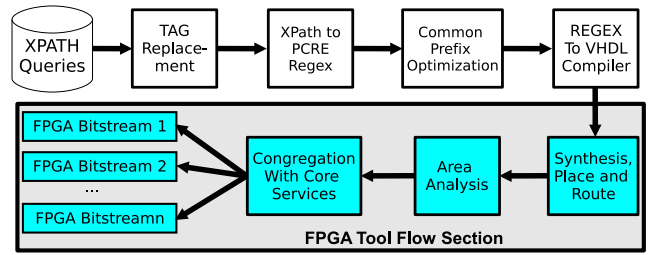

Figure 2: Compilation Flow of XPath expressions to FPGAs. The XPATH profiles go through a four step compilation process to generate the HDL. The lower gray section denotes the hardware flow for converting HDL to a bitstream for the FPGA.

As it can be seen from Figure 2 in the first step of the compilation workflow the tag elements in the XPath expressions, representing the user profiles, are replaced with fixed length string encodings. This is done to simplify the processing and to ensure that each tag element occupies the minimum amount of area possible on the FPGA device. Reducing the footprint of the individual XML tags results in higher query density on the chip and thus better usage of the hardware.

After this step the XPath expressions are translated to their equivalent PCRE form. During this translation process the navigation directions inside the XPath expression ( parent-child "/" and ancestor-descendant "//" ) are replaced with their PCRE counterparts. We describe this process in detail later in this section. In order to further reduce the query footprint on the FPGA device we cluster the regular expressions by their common prefixes. Those common prefixes are implemented as a single block on the FPGA unit. The result from the clustering step is a forest of "common prefix" trees. Each tree is compiled to generate a set of VHDL hardware blocks. The rest of the workflow involves FPGA specific compilation steps which will be discussed later as well.

### 3.1 Dictionary Replacement

The area of the FPGA chip is a limited resource. In order to get better usage, we minimize the tag footprint on the chip through a dictionary replacement process which replaces the XML tags in the input documents and the user profiles with fixed length strings. In

Table 1: PCRE operators used for parsing XML tags.

| Operator | Meaning |
|---|---|
| \w | Matches A to Z, a to z, 0-9, _ |
| \s | Matches a blank space |
| \c | Matches A to Z, a to z |
| \d | Matches a Decimal digit |
| + | Repeat 1 or more times |
| * | Repeat 0 or more times |
| \| | OR |

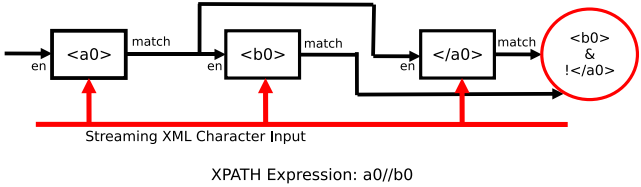

XPATH Expression: a0//b0

Figure 3: The block diagram for XPath <a0>//<b0>, showing the implementation of the ancestor-descendant axis

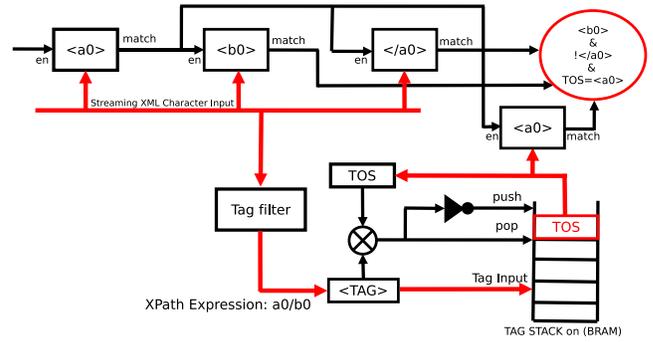

Figure 4: The block diagram for XPath <a0>/<b0>, showing the implementation of the parent-child axis. The additional hardware includes the tag filter, stack and TOS match blocks.

our implementation the length of the strings is set to 2 symbols which means that the size of all open tags is limited to 32 bits (2 symbols plus 2 tag markers of length 8 bits) and for close tags to 40 bits. As an example, the <test.document> tag is mapped to <a1>, while the closing tag </test.document> would map to </a1>.

### 3.2 XPath to Stack-enhanced Regular Expressions

If the XPath expression contains only the ancestor-descendant axis the translation to regular expression is straightforward. While the YFilter approach, maps an XPath profile to a sequence of NFA states connected with transitions, our approach maps an XPath profile to a regular expression. As an example the XPath profile "a0//b0" will be translated to the regex " <a0> [\w\s]$^+$ [<\c\d> | </\c\d>]$^*$ <b0> ". The various regular expression operators are explained in Table 1.

The regular expression in the above example accepts a sequence of XML tags which starts with <a0> and includes <b0>. It first matches the tag <a0>. Once this is matched, it will look for one (or more) characters (the [\w\s]$^+$ part) corresponding to text between tags and then will check for any number (0 or more) of open OR closed tags (the [<\c\d> | </\c\d>]$^*$ part) before it matches <b0>.

Moreover, in order for <b0> to be a descendant of <a0> in the document, the regular expression should match before the closing of <a0>. To implement this, during the hardware generation step for this regular expression, our compiler automatically adds a negation block on </a0> so that <b0> is matched before </a0> appears in the stream. The block diagram of the regular expression on the FPGA is shown in Figure 3. Each block represents a tag parser that searches for the given tag in the document stream. The right most hardware block (depicted as a circle), provides the final result from the matching process of the regular expression. Each block receives input from the 8 bit streaming XML interface and works in parallel with the other blocks.

The translation of the parent-child axis to a regular expression requires special treatment. This is due to the fact that the regular expressions are memoryless structures and one needs to ensure that the matched XML tags occur on consecutive levels in the document. For example, the level on which the parent is matched should be remembered so as to ensure that the child is matched on a consecutive level (e.g. it is immediately below the parent). The regular expression hardware is thus modified to include the notion of memory. In our implementation this is accomplished through the use of a tag stack which keeps the current path in the XML document. When an open tag is encountered it is pushed into the stack. Similarly when a close tag is reached it is popped from the top of the stack (TOS).

An example of a XPath expression that includes parent-child axis is shown in Figure 4. The XPath expression "a0/b0" is translated to a modified regular expression with a stack control directive. The modified regular expression is: " <a0> [\w\s]$^+$ [<\c\d> | </\c\d>]$^*$ [Stack1] <b0> ".

When testing a parent-child relationship, in addition to checking for the ancestor-descendant property we have to ensure that the level difference between the respective tags is one. Hence we use an extra hardware block – the TOS matching –, which continuously monitors the top of the stack and ascertains that the matched element <b0> is indeed a child of the previously matched element <a0>.

Figure 4 describes how we monitor the current level. The XML tag stack block, works in parallel with the ancestor-descendant block on the FPGA. The additional Tag Filter block extracts XML tags from the document stream. When an open XML tag is extracted, it triggers the push function and this tag gets pushed into the stack. In a similar way closing tags trigger the pop function and remove the head of the stack. A difference with the previous ancestor-descendant match is that finding <b0> after <a0> is not enough; we need also that the top of the stack is <a0> (when <b0> is found). Since many regular expressions are using the same XML input data stream, we need only one stack block per data stream.

### 3.3 Common Prefix Optimization

The regular expressions derived from the XPath profiles typically depict large commonality in their prefixes. For example "a0//b0//c0//d0" and "a0//b0//c0//e0" share the common prefix "a0//b0//c0", with corresponding suffixes "d0" and "e0". The hardware cost of implementing the regular expressions is measured in terms of the FPGA area used to implement the logic. It is thus advantageous to combine multiple regular expressions into a common prefix tree. Such a tree can help reduce the area cost of the hardware by implementing the common prefix as a single block on the chip. In the above example, instead of implementing two regular expression hardware blocks and duplicating the "a0//b0//c0" logic, we can have a single implementation for the common path. As a result, more profiles can fit in a given FPGA area.

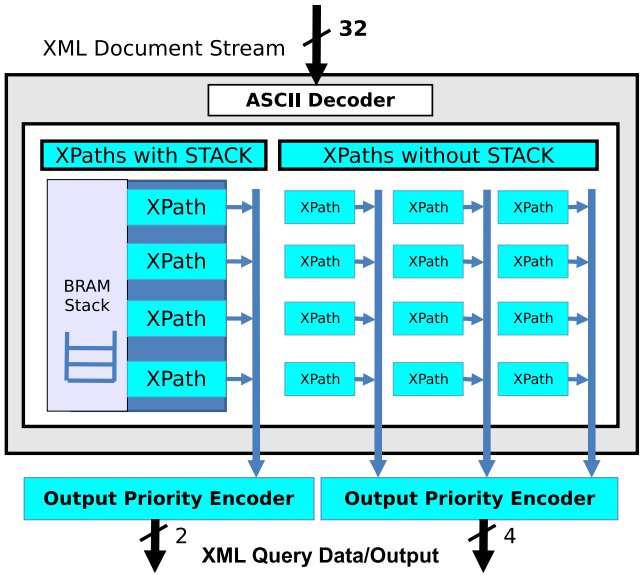

Figure 5: An example FPGA organization denoting the input / output data path with sixteen XPath expressions

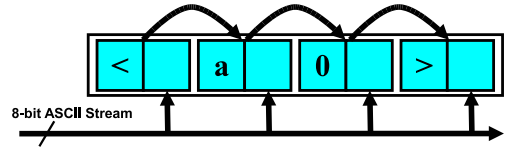

Figure 6: Block diagram of the Character Match Hardware Block for a tag <a0>. The hardware is a 8-bit x 4 comparator block.

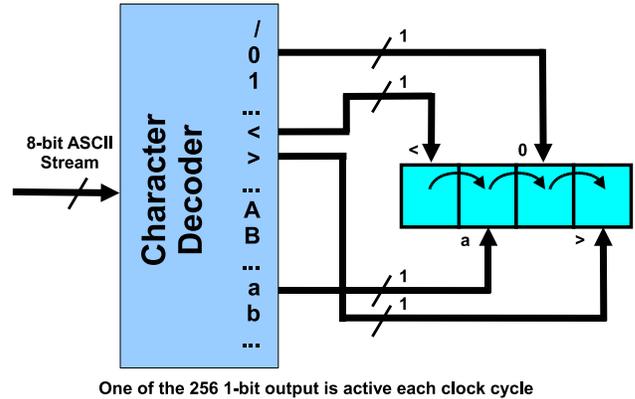

Figure 7: Block diagram of the Character Pre-Decoder Hardware Block for a tag <a0>. The hardware is a 1-bit x 4 comparator block.

Given a set of XPath profiles, we first create their regular expressions and then sort them in alphabetical order. We then run a common prefix discovery algorithm on the sorted list of the regular expressions. The algorithm recursively grows the common prefix one tag at a time. The result is a forest of common prefix trees, each representing a set of profiles. From these trees we then create the FPGA hardware.

### 3.4 Area Efficient Character Decoder Hardware

Implementing XPath profiles on FPGAs mainly involves implementing character matching blocks to identify XML tags in the input document stream. The character matching hardware block [9] compares sequences of characters from the input stream to a given sequence that defines an XML tag. Figure 6 exemplifies the comparator hardware that matches an XML tag. Each character requires an 8-bit comparator block. The implemented character matching blocks for the XML tags consist of many redundant blocks, the prime examples being the open "<", close ">", and end tag "/" characters.

It is possible to simplify the character match hardware with a 8-bit ASCII character pre-decoder. The character pre-decoding hardware decodes the incoming ASCII data stream at the input. An 8-bit input is decoded into one of 256 possible 1-bit character signal every clock cycle. As an example, if the input was HEX "0x60", the output line for the character "a" would be high on that clock cycle and the rest of the other 255 outputs would be all zeros. The character decoder hardware block simplifies character matching by replacing 8-bit character match hardware blocks with a 1-bit comparator and results in area efficient hardware. Figure 7 depicts the character pre-decoder block, and the simplified 1-bit comparator blocks for matching an XML tag. Moreover since 1-bit data lines are routed on the FPGA for each character in the XML tag, the FPGA routing overhead is reduced, which in turn leads to a design which offers faster clock speed.

### 3.5 FPGA Implementation of Regular Expressions

A regular expression syntax could be defined using various syntaxes such as PERL, UNIX, etc. Our implementation uses the PERL semantics. The compiler uses a modified version of the PCRE library v6.7 compilation flow. It simulates the behavior of a regular expression in VHDL, suitable for implementation on FPGA. We modified the compiler to take into account the stack directives and generate the hardware blocks to support parent-child axes.

After obtaining the VHDL sources for the user profiles, we generate additional hardware blocks including an input ASCII decoder, two output priority encoders (one each for queries with or without parent-child axes) and the tag stack. We group the VHDL sources into two sets, i.e. profiles without parent-child axes and profiles with parent-child axes. The organization of XPath expressions on the FPGA is depicted with an example in Figure 5. The four XPath profiles on the left correspond to expressions that contain parent-child axes and thus use the on-chip FPGA stack. When the streaming document matches a given profile, the output priority encoder is set to that profile.

We synthesize the generated VHDL code, using the XILINX synthesis tool to obtain the hardware netlist. The next step involves running the Place and Route tool, which report the clock frequency of the hardware design.

Our target FPGA is a Virtex-4 LX 200 device, and the target hardware is the Silicon Graphics RASC RC 100 board. In order for our FPGAs to run on this board we had to add a hardware module (RASC Core Services) which allows us to send and receive data and control the FPGA from the host system. Finally we generate the bitstreams that are loaded on the FPGA.

## 4. EXPERIMENTAL EVALUATION

This section describes our experimental setup and the obtained results when comparing the throughput of XML filtering performed on FPGAs, with respect to software based filtering solutions, i.e. the YFilter. This system is widely adopted as a software-based XML filtering approach. The software part of the experimental evaluation was executed on a Core 2 Quad 2.66 GHz with 8GB of RAM available. We choose YFilter because it uses more general approach for the XML filtering compared to other existing solutions. For example the lazy DFA presented in [12] has been shown to provide faster performance than the YFilter, but nevertheless assumes certain constraints for documents and profiles. We leave comparisons with such systems for future work.

In order to provide in depth evaluation of the performance for both the hardware and software implementations, we use the *ToXGene* XML document generator [7]. This tool generates XPath profile datasets for a specified DTD structure. We use the same set of profiles to test all methods described in this section.

We have generated multiple sets of profiles with varying path length, i.e. 2 Tags, 4 Tags and 6 Tags using the PathGenerator class in YFilter. The number of queries in each set varies from 16 to 1024. The streaming documents, used in the evaluation, vary in size from one to eight MBs.

During the experimental evaluation of the software approach we measure the throughput of the system (the size of the document set in megabytes provided as input divided by the time in seconds between the moment when the set enters the system to the moment those documents are filtered by the matching process).

For the hardware implementation we use the Silicon Graphics Altix 4700 [5] supercomputer system along with the RASC RC 100 [35] blade. We stream XML data stored in the memory (RAM) of the Altix system to the FPGAs placed on the RASC blade. We also stream the output of the priority encoders from the FPGA back to the Altix system. The output of the priority encoders is also continuously decoded by the host system, to filter the XPath expressions that have a match with the current document. As an output we provide the profile that is successfully matched as well the location of the match inside the document structure.

The speed in the hardware implementation is also measured in terms of throughput (MBytes/s). However we also measure the area occupied by the hardware design since it is considered a critical resource for FPGAs. In order to obtain a better understanding of the area/speed tradeoff which is something typical of FPGA based systems, we progressively increase the number of XPaths profiles processed on the FPGA. The total number varies from 16 up to 1024 profiles.

### 4.1 Area Utilization

With the first set of experiments we identify the impact of our two optimizations (i) the common prefix and the (ii) character pre-decoder on the area occupied on the chip. We consider four implementation scenarios:

- Unoptimized Hardware (**Unop**): A system implementation without character decoding and with no common prefix optimization.

- Common Prefix Optimized Hardware (**Com-P**): A system which uses the common prefix optimization of the queries but without character pre-decoding.

- Unoptimized Hardware with Character Decoding (**Unop-CharDec**): A system that utilizes a character pre-decoding blocks, but without common prefix optimization.

- Common Prefix Optimized Hardware with Character Decoding (**Com-P-CharDec**): A system which takes advantage of both optimizations.

The results from these experiments are shown on Figure 8. The general trend which can be observed across the plots is that the occupied area increases linearly with increasing number of XPath queries for a given XPath length. The increased length of the queries have the same impact over the area. As expected the unoptimized hardware implementation is the one which consumes most area out of all implementational scenarios. Sometimes this can be prohibitively expensive. For example we were unable to implement the dataset which contains 1024 Xpath Queries with 6 tags with this approach because of space limitation on our FPGA.

In contrast the implementation which uses the common prefix optimization along with character decoder produces the most efficient area implementation of Xpath profiles. This approach is highly efficient when compared to the unoptimized hardware and in most cases provides five to seven times area improvement.

### 4.2 Performance Speedup

In this experimental set we compare the performance of both the hardware and software approaches. We use the same implementational scenarios discussed above with the same set of queries. The results of the comparison can be depicted in Figure 9.

In particular, there is a gradual reduction in throughput with the increase of the number of XPath profiles implemented on the FPGA. On average the unoptimized character pre-decoder based FPGA design for XPath filtering offers higher throughput than other designs. The design that almost always leads to the slowest speeds is the hardware implementation of common prefix optimized regular expressions without the use of character pre-decoder hardware.

Here we also compare with the performance of the software approach (YFilter). A common characteristic is that all FPGA based solutions are orders of magnitude (at cases 100 times) faster than YFilter. The performance of YFilter appears constant because it is limited from above by the bottleneck which appears in the traditional von Neumann architectures.

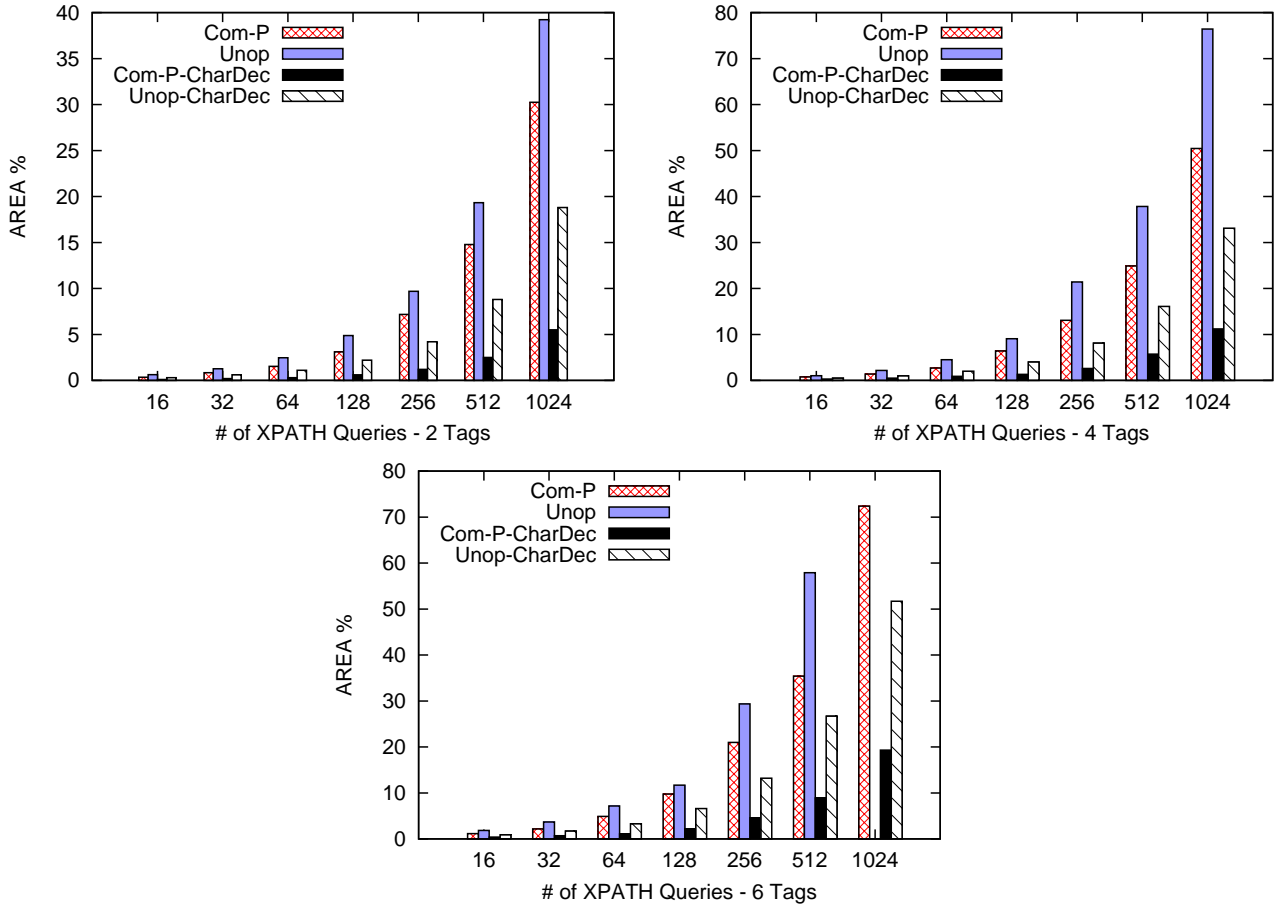

**Figure 8: Variation of FPGA Area (in %) with increasing number of XPath expressions**

### 4.3 Summary

The area/speed tradeoff, typical for FPGAa, is apparent in the scenario with the implementation that uses the common prefix optimization with character pre-decoder. This approach provides the maximum area efficiency but with low throughput which for most cases is almost as low as common prefix optimized hardware. In general the common prefix optimization, even though improves area efficiency, brings down the clock speed, and thus the throughput of a FPGA based hardware filtering design. The second optimization that uses character pre-decoders offers better area/speed tradeoff.

The overall results of our experimentation lead to the conclusion that using an FPGA for parallel and efficient XPath filtering approach provides orders of magnitude throughput improvement (around 100 times for some datasets). It was observed that increasing XPath lengths decreased the speedup offered by FPGA. The same is also true about the increasing number of XPath profiles implemented on FPGA. Moreover common prefix optimized hardware, both with and without character pre-decoder provides better area utilization but lowers the system throughput. The reason is that, adding hardware complexity leads to lower clock rates on the FPGA. The unoptimized character decoder based FPGA implementation of XPath filtering offers the best area speed tradeoff.

## 5. CONCLUSIONS AND OPEN PROBLEMS

This paper provides a preliminary implementation of XML filtering using a flexible FPGA architecture. Such filtering is limited in traditional Von Neumann architectures by the presence of a bottleneck between the CPU and memory. The execution of a single instruction requires multiple clock cycles for fetching, processing and storing the data back into memory. Using FPGAs alleviates this problem, by removing unnecessary operations and performing an instruction over the streaming data in a single clock cycle. Our experimental evaluation reveals order of magnitude (around 100 times) improvement in the performance speedup. We presented a hardware design and optimizations for efficiently handling XPath profile queries.

The idea of combining FPGAs and XML processing leads to many directions for further research. While our approach processes documents in a 'top-down' fashion, it is interesting to examine whether a 'bottom-up' solution is also possible. This means that document paths are first stacked and their leaf nodes are examined first for a match (which will be advantageous for documents whose more selective tags are at the leaves). A comparison with other software implemented XML filtering techniques (like lazy-DFA based approaches) is also interesting. One open problem is how to deal with dynamic updates (deletions and insertions) on the profile queries.

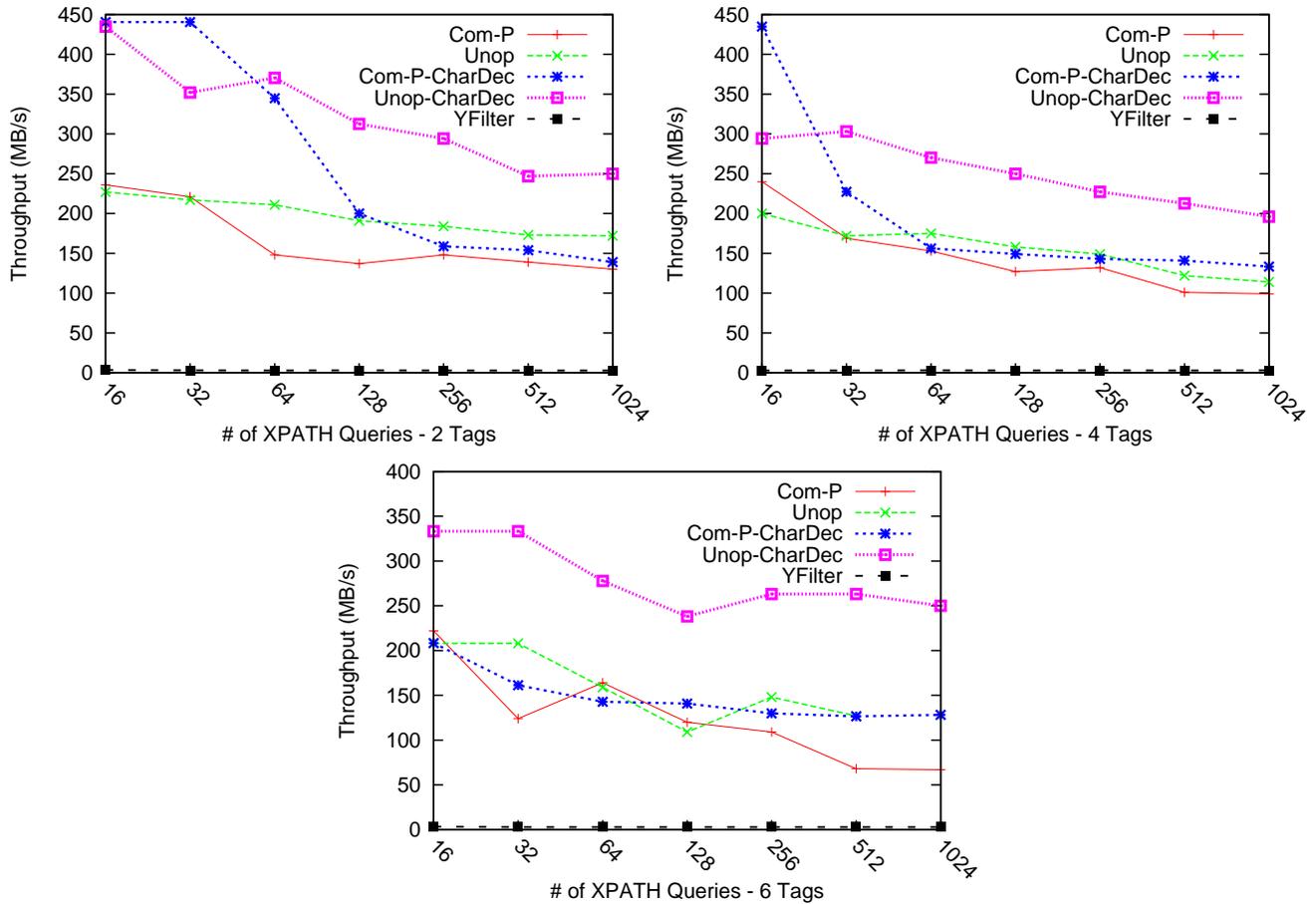

Figure 9: FPGA and YFilter (Software) throughput comparison with increasing number of XPath profiles.

A natural extension is to provide support for twig profiles. Unlike XPath profiles, which can only look for the presence of a given path inside the structure of the XML document, a twig pattern query identifies more complex structures like trees. To support twig profiles in our system we need a different approach that the architecture described in Section 3.

A straightforward solution for the twig pattern matching problem is to decompose the twig query into individual paths and process each path separately. The results from the individual paths are then joined together in a post processing step to produce the final outcome of the query. This approach (which would also work with our current XPath architecture) however requires extra processing time: first, there may be many path matches not related to the twig (false positives that need to be eliminated); second, the common sections of individual paths will be processed multiple times which is redundant.

Instead, a more promising approach is to employ *holistic* twig profile filtering, based on the Prüfer sequence [29] encodings of the XML document and the profiles. A Prüfer sequence was originally used in graph theory to describe a *unique sequential* encoding of a labeled tree. Since both the streaming XML document and the profiles represent trees, such encoding is easily attainable through tree traversals. This approach has been used in the past in software based XML filtering systems like PRIX [30] and FiST [16].

The main idea is to reduce the problem of twig matching to subsequence matching between the document and the profiles.

The reason why the Prüfer encoding of XML documents is an appealing method for identifying twig pattern matches within a document is because it captures well the document's structure. In particular, the sequence carries enough information to check parent-child and ancestor-descendant relationships within a tree structure. In particular, if a tree Q is a subgraph of another tree T then the Prüfer encoding of Q is a subsequence of the Prüfer encoding of T [30]. The reverse however is not true (i.e., we can have false positives). Since the nature of subsequence matching leads also to sequential processing over the document (like parsing and filtering), we can then take advantage of the FPGA properties. We have currently an initial implementation of an FPGA architecture for supporting twig matching and are experimenting with approaches to efficiently eliminate false positives within the FPGA.

Another interesting future direction would be a comparison between multiple FPGAs and multi-core machines. Finally, an orthogonal problem is whether FPGAs can enable faster multi-query XML processing over an archived collection of documents.

**Acknowledgements:** This research was partially supported by NSF grants CCF-0811416 and IIS-0705916 as well a gift from CISCO. Marcos R. Vieira's work has been funded by a CAPES (Brazilian Federal Agency for Post-Graduate Education)/Fulbright Ph.D. fellowship.